# Wafer scale reactive sputtering of highly oriented and ferroelectric $Al_{0.6}Sc_{0.4}N$ from 300 mm AlSc Targets

Tom-Niklas Kreutzer[1]*, Muhammad Zubair Ghori[1], Md Redwanul Islam[2], Fabian Lofink[1,2], Fabian Stoppel[1], Axel Müller-Groeling[2,3], Simon Fichtner[1,2]

(1) Fraunhofer Institute for Silicon Technology (ISIT), Fraunhoferstrasse 1, 25524 Itzehoe, Germany
(2) Kiel University, Institute for Material Science, Kaiserstrasse 2, 24143 Kiel, Germany
(3) Fraunhofer-Gesellschaft zur Förderung der angewandten Forschung e.V., Postfach 200733, 80007 Munich, Germany
*Corresponding author: Tom-Niklas.Kreutzer@isit.fraunhofer.de

**Abstract**

This paper presents progress towards the large-scale manufacturability of piezo- and ferroelectric $Al_{1-x}Sc_xN$ thin films with very high Sc content. $Al_{0.6}Sc_{0.4}N$ layers were deposited by pulsed DC reactive sputtering from a 300 mm diameter $Al_{0.6}Sc_{0.4}$ target on standard 200 mm Si wafers with Pt bottom- and Mo top-electrodes. The deposited films were analyzed in depth with X-Ray diffraction (XRD), Reciprocal Space Mapping (RSM), Scanning electron microscopy (SEM) and Energy Dispersive X-Ray Spectroscopy (EDX) showing well oriented c-axis growth over the full wafer with slight variation in the film thickness and Sc content over the wafer radius. An overall low density of abnormally oriented grains (AOG) was found. Further wafer mapping for piezoelectric and dielectric properties showed a piezoelectric performance increase by 40 % in comparison to $Al_{0.7}Sc_{0.3}N$ while only moderately increasing the permittivity and loss factor. Switching measurements revealed ferroelectric behavior of the film on all measured positions with an average remanent polarization of 88.36 µC/cm$^2$ and an average coercive field of 244 V/µm. This successful demonstration opens new opportunities for MEMS applications with demands for high forces like microspeakers or quasi static micromirrors.

**Keywords**

AlN, AlScN, ScAlN, MEMS, Piezoelectric, Ferroelectric, Sputter Deposition, Scandium, Wafer Scale

## Introduction

Immediately after the discovery of the enhanced piezoelectric activity of ScN alloyed AlN, a significant scientific and commercial interest in $Al_{1-x}Sc_xN$ arose [1]. Within a few years after its initial discovery, first bulk-acoustic-wave (BAW) devices based on the compound entered the market [2]. Today material integration in other devices like micro-electro-mechanical-systems (MEMS) actuators and, thanks to its ferroelectric properties, also ferroelectric memories, is pursued [3–5].

The widespread interest in $Al_{1-x}Sc_xN$ and related materials can be understood by the clear advantages the wurzite nitrides like AlN and $Al_{1-x}Sc_xN$ possess in comparison to common perovskite oxides like Lead-Zirconate-Titanate (PZT). They allow for straight forward post - Complementary Metal-Oxide Semiconductor (CMOS) compatible integration, avoid Restriction of Hazardous Substances Directive (RoHS) critical lead, while also offering lower permittivity and loss factors. The only downside of, especially AlN, in comparison to PZT is its vastly lower piezoelectric coefficient. ScN doping increases the transverse thin film coefficient $e_{31,f}$ by more than 300 % in comparison to pure AlN [6] and thus allows for higher driving forces in MEMS drives and increased coupling in surface-acoustic-wave (SAW) and BAW devices [7–9].

Since the piezoelectric coefficients increase monotonically with ScN content up to about 43 % ScN, many applications aim for the highest possible ScN content. These high ScN concentrations, especially near the wurtzite-to-cubic phase transition at 43 % [1, 10], cause a variety of challenges when upscaling to industrial standards: Most crucially, difficulties in target manufacturing and insufficiently polar crystalline orientation have so far limited large scale applications to around and below 30 % ScN. Well controlled $Al_{1-x}Sc_xN$ growth with ScN concentrations around 7 % [11] and near 30 % were previously demonstrated on standard 200 mm wafers. Lin et al. [12] grew $Al_{0.7}Sc_{0.3}N$ film with thicknesses from 100 nm to 1 µm on Mo electrodes via pulsed DC magnetron sputtering from an AlSc alloy target on 200 mm wafer. Mertin, Nyffeler et al. [13] sputter deposited 1 µm $Al_{0.67}Sc_{0.33}N$ films on Pt and Mo electrodes from an Alloy target on 200 mm wafer. Furthermore, Mertin, Heinz et al. [14] grew 1 µm $Al_{0.67}Sc_{0.33}N$ via pulsed DC magnetron sputtering on Pt electrodes from an $Al_{0.7}Sc_{0.3}$ target on 200 mm wafer and Barth, Schreiber et al. [15] deposited $Al_{0.695}Sc_{0.295}N$ up to several micrometer on Pt electrodes.

Here, we demonstrate that also $Al_{0.6}Sc_{0.4}N$ can be grown with decent homogeneity over 200 mm wafers from 300 mm targets using an industrial sputter tool. The size of the target is a key factor in the successful growth, as the large diameter is essential for a mass-production-suitable homogeneity and deposition rate. Previously no large targets, especially at these extremely high Scandium contents, were available due to manufacturing issues arising from intermetallic phases and the brittleness of the alloy [16, 17]. Moreover, we demonstrate that, using Pt as a bottom electrode, $Al_{0.6}Sc_{0.4}N$ layers with little to no abnormally oriented grains (AOGs) over the majority of the wafer can be deposited on 200 mm diameter substrates. This presents a significant step forward, towards the general availability of high-performance piezoelectric films for the commercial production of demanding MEMS-actuator applications using $Al_{1-x}Sc_xN$.

**Method**

The film stack was deposited on standard 200 mm Si-(100) wafers with a thickness of 725 µm, fully covered by a wet oxide of 650 nm thickness on both sides. The complete stack consists of 22 ± 0.8 nm Ti as an adhesion promoter, a 96 ± 2 nm Pt bottom electrode for ideal growth conditions, followed by 595 ± 15 nm $Al_{0.6}Sc_{0.4}N$ and covered by a 214 ± 3 nm thick Mo top electrode for ease of structuring. The thickness and tolerance for all metallic films arises from process qualifications via four-point resistance measurement and were not directly measured on the substrates shown in the following. Thickness and variation for the $Al_{0.6}Sc_{0.4}N$ film was collected via SEM cross-sections as described in the results section. All deposition steps were performed in an Evatec Clusterline 200. The metallic films were DC sputtered with Ar as the plasma species, Ti and Pt in a multi-source chamber with rotatable shutter while Mo was deposited in a single chamber. Ti deposition took 50 s, followed by 115 s of Pt sputtering. The final Mo deposition time was 95 s. The $Al_{0.6}Sc_{0.4}N$ film was sputtered from a 300 mm diameter $Al_{0.6}Sc_{0.4}$ alloy target, manufactured by Materion. The piezoelectric layer was grown via pulsed DC reactive sputtering with a deposition rate of 3.3 µm/h. Resulting in a deposition time of 650 s for the $Al_{0.6}Sc_{0.4}N$ film. Nitrogen and Argon were used as the plasma species. The Nitrogen is used in the reaction of AlSc to $Al_{1-x}Sc_xN$ at the target surface. The nitrogen flow is tightly controlled via a mass-flow-controller (MFC) to avoid metallic sputtering and target poisoning on the other end [18]. All process gases are of 5N purity. The $Al_{0.6}Sc_{0.4}N$ film stress was tuned via RF biasing of the wafer chuck. This method allows for precise stress control over a large range of possible stress values. Before and after the full stack deposition, the wafer bow was measured via capacitive sensing with an Endress+Hauser MX 203, this allowed for wafer stress calculation via the Stoney equation [19, 20]. The film growth was evaluated via scanning electron microscope (SEM) in a Hitachi S-9260A. SEM imaging allows for quick detection of AOGs [21].

To continue the $Al_{1-x}Sc_xN$ characterization, the material stack was processed into capacitor test structures. Firstly, the Mo top electrode was opened in small areas with standard phosphoric-acid wet etchant PWS 80-16-4(65) by Honeywell (PWS), subsequently acting as a hard mask. With this hard mask the $Al_{1-x}Sc_xN$ was opened to the Pt bottom electrode in a Tetramethylammonium hydroxide (TMAH) wet etching bath. Finally, the Mo top electrode was pattered to its final square capacitor shapes with a second lithography sequence and again PWS etchant. By this method a continuous bottom electrode and near continuous $Al_{1-x}Sc_xN$ layer remain on the wafer and thus reduce processing impacts on the test structures.

The finished wafer was then characterized for its piezo-, ferro- and dielectric properties with a double-beam-laser-interferometer (DBLI) manufactured by aixACCT and fitted with a TF2000 analyzer. Square capacitor structures with side-lengths of 1 mm and 0.5 mm were characterized. As these sizes do not match to the substrate thickness, both were measured via the DBLI with the same parameters (PZM) and a linear interpolation was used to calculate the correct out-of-plane effective piezoelectric coefficient $d_{33,f}$ Klicken oder tippen Sie hier, um Text einzugeben.[6, 22–24]. PZM measurements were performed with 100 V amplitude at 200 Hz triangle signal with 999 averages per test structure. Polarization values are calculated by integration of the UI curve. On the 0.5 mm pads CV-curves (CVM) were recorded to determine the materials relative permittivity $\varepsilon_r$ and the loss factor *tan(δ)* in percent. The CV curves were recorded with a small signal frequency of 5 kHz and an amplitude of 5 V. Furthermore, a DC breakdown measurement (BDM) for film leakage and leakage compensated ferroelectric PE loops were recorded via the positive-up-negative-down (PUND) scheme [25, 26]. The PUND measurement was done on seven positions over the wafer. In total the measurement scheme adhered to the following sequence: For piezoelectric characterization: I). PZM mapping, II.) CVM mapping. For ferroelectric characterization (starting on fresh capacitors): Initial PZM, 3000 major cycles to remove imprint, PZM, PUND and finally BDM.

In order to investigate possible correlations between the (piezo-) electrical properties and the crystalline structure, XRD measurements were conducted on different positions of the wafer. In a Malvern Panalytical X'Pert³ MRD XL θ-2θ scans from 20° to 90° and rocking curves of the $Al_{0.6}Sc_{0.4}N$ (002) and Pt (111) reflections were recorded on seven positions on the wafer. The tool was fitted with a copper anode tube in combination with a Ni-Kα-filter.

Each step-field of 25 by 25 mm was cut into seven dies, six 25 by 3 mm beams and one 25 by 7 mm piece. On these larger pieces, reciprocal space maps (RSM) around the $Al_{0.6}Sc_{0.4}N$ (002) reflex were collected in a Rigaku SmartLab on three positions. On the same positions additional asymmetric scans of the $Al_{0.6}Sc_{0.4}N$ (105) reflex allowed the determination of the a and c lattice parameter of the $Al_{0.6}Sc_{0.4}N$ film.

The chemical composition and film thickness were acquired with the help of a Zeiss Ultra plus 55 SEM fitted with an Oxford Instruments Ultim Max 65 Energy-Dispersive-X-Ray-Spectroscopy (EDX) detector. This SEM was also used for high resolution images of the $Al_{0.6}Sc_{0.4}N$ surface.

## Results

The test design was stepped in a seven-by-seven grid on the wafer. Thus, all measurements follow this grid pattern on the wafer. The (piezo-) electric characterization via DBLI mapped the whole grid, while the structural characterization was performed on a selection of dies arranged in L-shape as shown in Figure 1. Subsequent graphs refer to the positions in this figure.

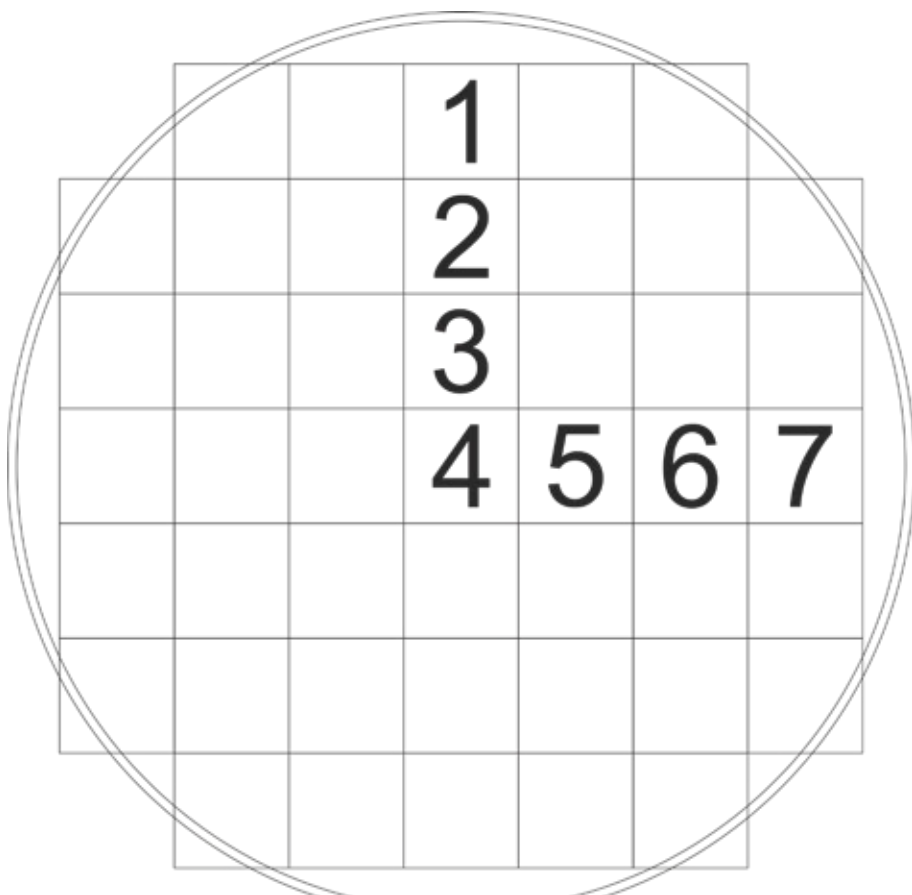

*Figure 1: Schematic of wafer stepping. All measurements were performed on the numbered dies and this numbering is used to correlate results with the position on the wafer. In addition, (piezo-) electrical characterization was conducted on the whole wafer grid.*

The SEM images of the wafer surface after full stack deposition are shown in Figure 2. These images were taken at equidistant positions over the whole wafer surface and assembled as a wafer map to give an impression of the stack morphology on the whole substrate. Each of the 13 images were taken with the same magnification such that the shown scalebar is valid for all separate images in the map. Overall, little to no AOGs were observed over the whole wafer, giving raise to the expectation of good piezoelectric properties due to the dominance of well c-oriented grains. A radius dependent change in the grain shape and size can be observed in the map. A coarser and apparently less dense surface texture is visible in the wafer center in comparison to the wafer edge. The change in texture does not seem to correlate with the density of AOGs in both texture regions. High resolution SEM images of the seven positions mentioned in Figure 1 are available in the supplemental material.

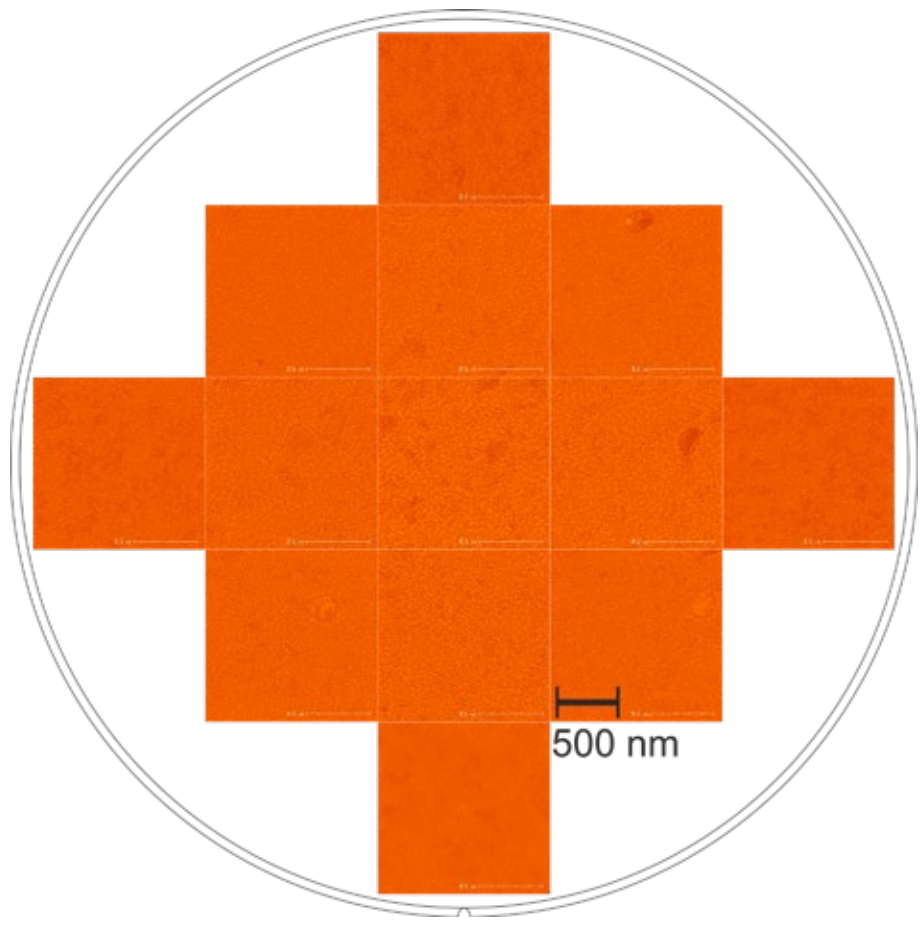

*Figure 2: SEM map after deposition of the full stack (Ti/Pt/$Al_{0.6}Sc_{0.4}N$/Mo). The separated images are arranged in a wafer map to indicate their position on the substrate, but do not show a connected image of the wafer surface. Overall, a low density of misaligned grain growth is observed, while a difference in grain shape can be observed, depending on the radius.*

In addition to SEM imaging of the films surface Focused Ion Beam (FIB) cuts were employed to reveal the cross-sectional structure, as shown in Figure 3.

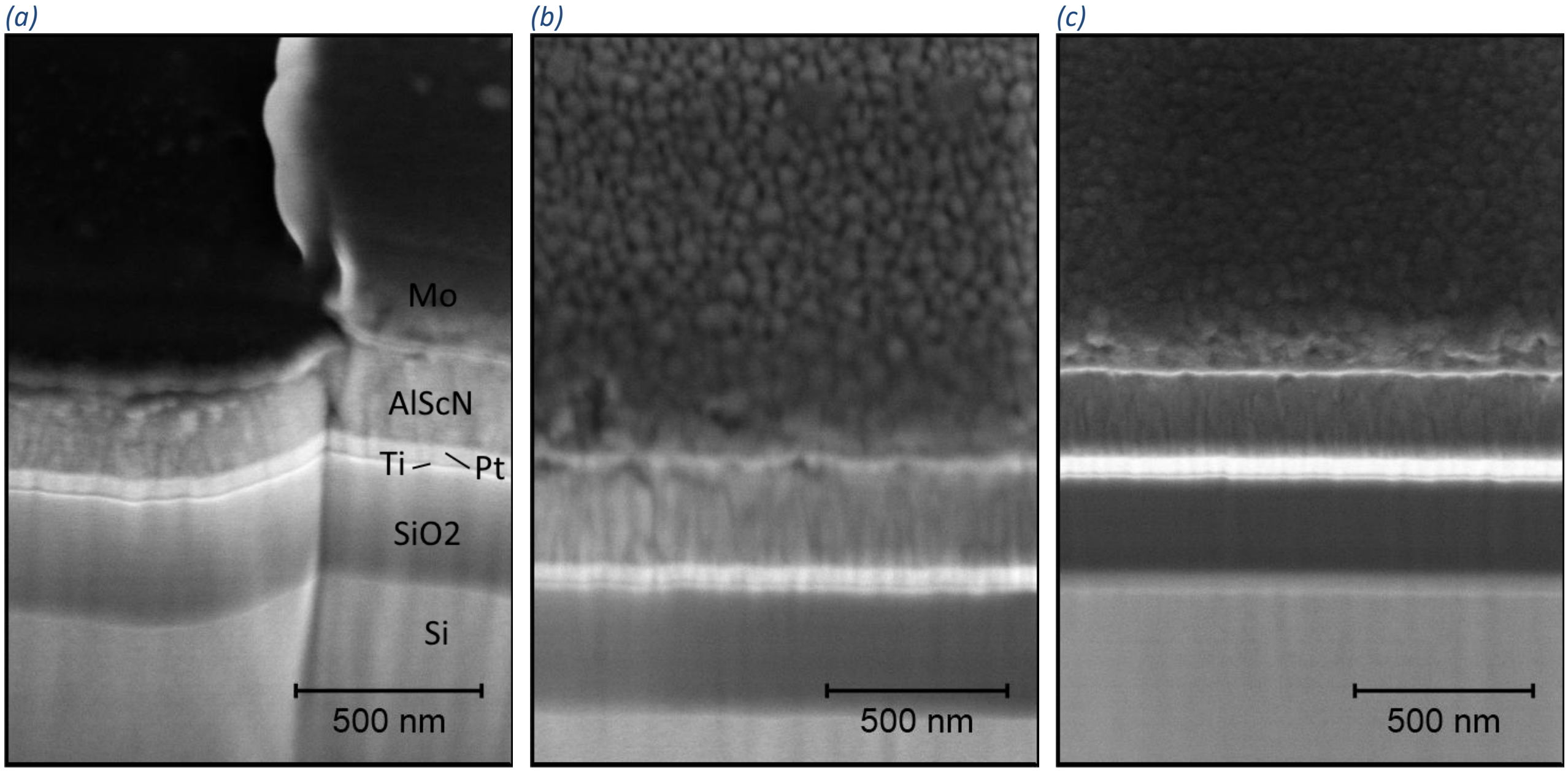

*Figure 3: FIB cross-sections imaged via SEM. A) At Position 1, the whole stack is visible, while the Mo top electrode is partly removed. The $Al_{0.6}Sc_{0.4}N$ film grew in a very dense film. B) Position 4, the columnar structure of the $Al_{0.6}Sc_{0.4}N$ is visible. C) Position 7, the densely $Al_{0.6}Sc_{0.4}N$ layer is comparable to position 1.*

These cross-sections also hint to a variation in crystal quality from edge to wafer center.

This change in growth could origin from different growth conditions or a local variation in the film composition. For a deeper look into the film growth and texture, X-ray diffraction (XRD) in form of θ-2θ scans (T2T), rocking curves (RC) and reciprocal space maps (RSM) were recorded.

The resulting T2T measurements from all seven positions on the wafer are shown in Figure 4. The position of the Pt bottom electrode reflections at 39.86° 2θ and 85.96° 2θ are virtually constant over the whole wafer, whereas the $Al_{0.6}Sc_{0.4}N$ (002) reflection at around 36.6° 2θ and the $Al_{0.6}Sc_{0.4}N$ (004) reflection at around 78° 2θ shift to lower angles towards the wafer center.

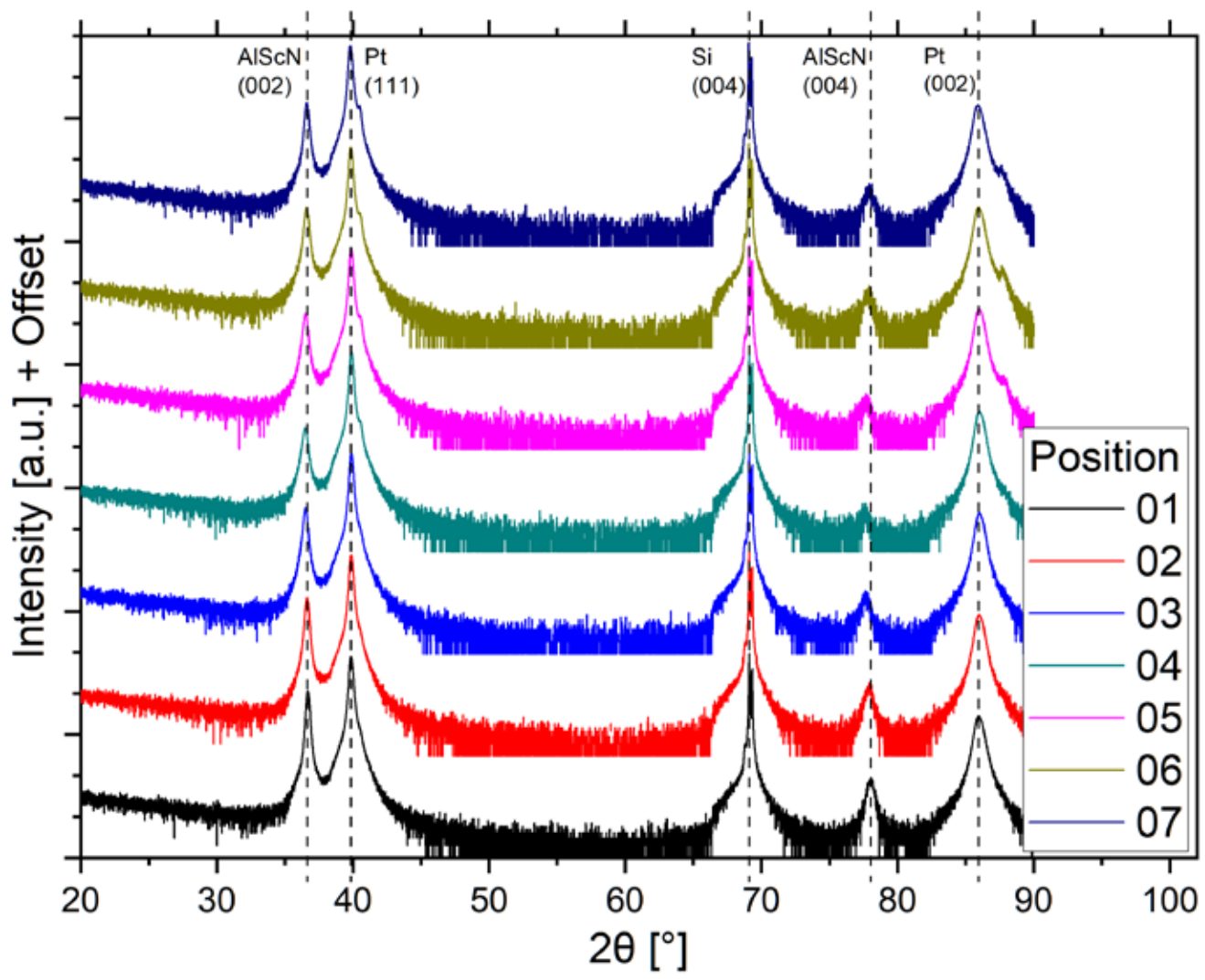

*Figure 4:Logarithmically plotted $\theta$-$2\theta$ scans for all seven dies. Measured at a position where the Mo top electrode was removed during processing. A wafer position dependency of the $Al_{0.6}Sc_{0.4}N$ (002) and (004) reflex can be observed as, in the wafer middle, the reflex shifts to lower angles.*

For the Pt (111) and the $Al_{0.6}Sc_{0.4}N$ (002) reflections, RCs were recorded. The Pt RCs, shown in Figure 5, are again confirmed to be virtually constant in their position and full width half maximum (FWHM) over all recorded positions. The RCs are centered around 20° ω with an average FWHM of 2.28°.

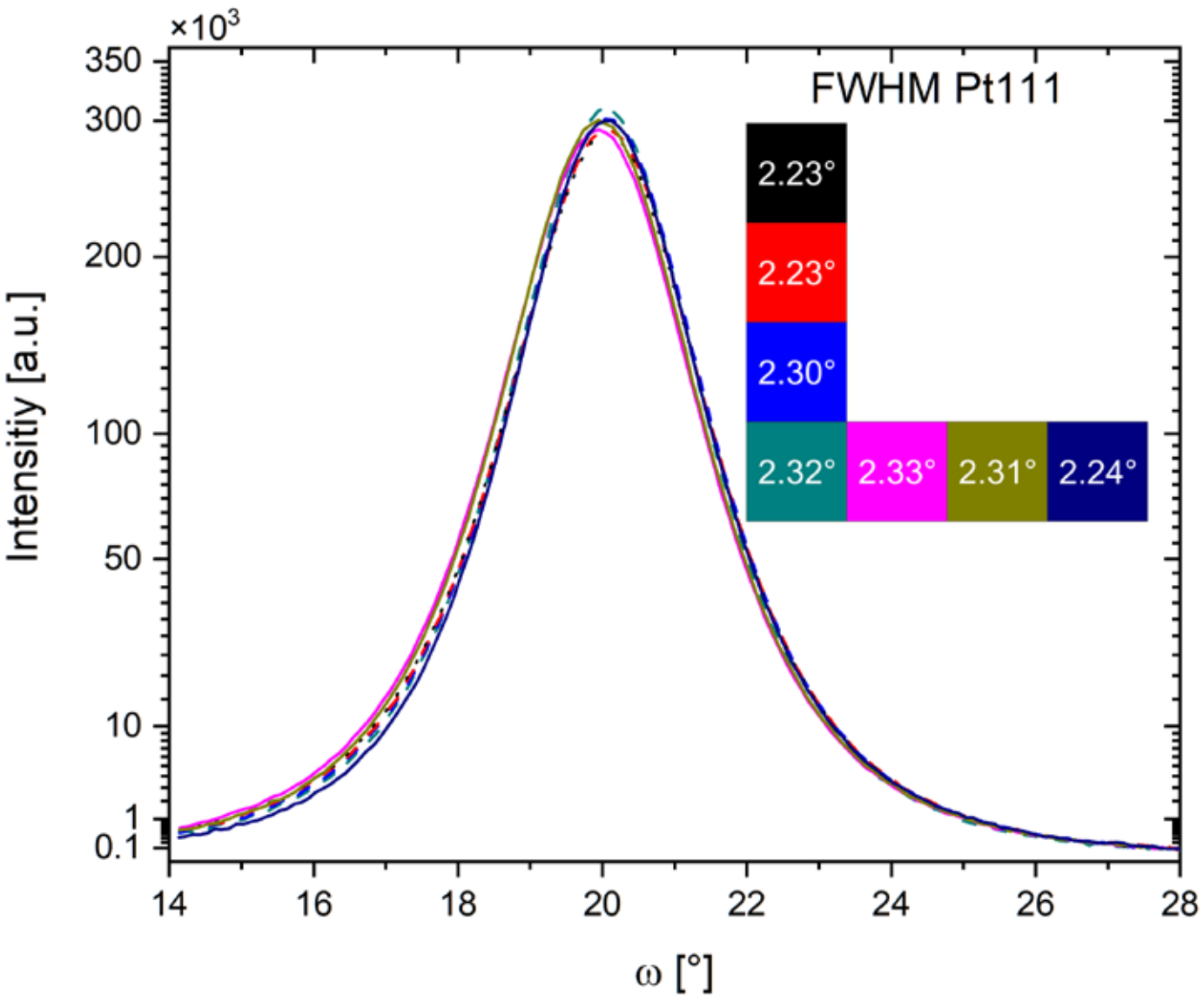

*Figure 5: Rocking curves of the Pt (111) reflex of the bottom electrode. The extracted FWHM is shown in the inset in the top right. The reflex position, width and intensity are stable over the whole wafer.*

The RCs of the $Al_{0.6}Sc_{0.4}N$ (002) reflection, as presented in Figure 6, differ more substantially in their width, position, and height in dependence of the position on the wafer. Two groups of RCs can be identified. The lowest intensity and largest FWHM is measured in the wafer center. The two groups correlate to the two measurement directions on the wafer. The group combining the measurement positions four to seven shares a common angle at maximum intensity, while the angle at maximum intensity in the group of the positions one to four drifts towards a higher center angle nearer to the wafer center. From center to edge the intensity increases and the FWHM decreases steadily from 5.01° down to 1.40° - substantially lower than the FWHM of the bottom electrode. This confirms that $Al_{0.6}Sc_{0.4}N$ is grown with very good texture over most of the wafer surface. More extensive fine-tuning of the process should furthermore result in a homogeneous texture also in the central die on the wafer.

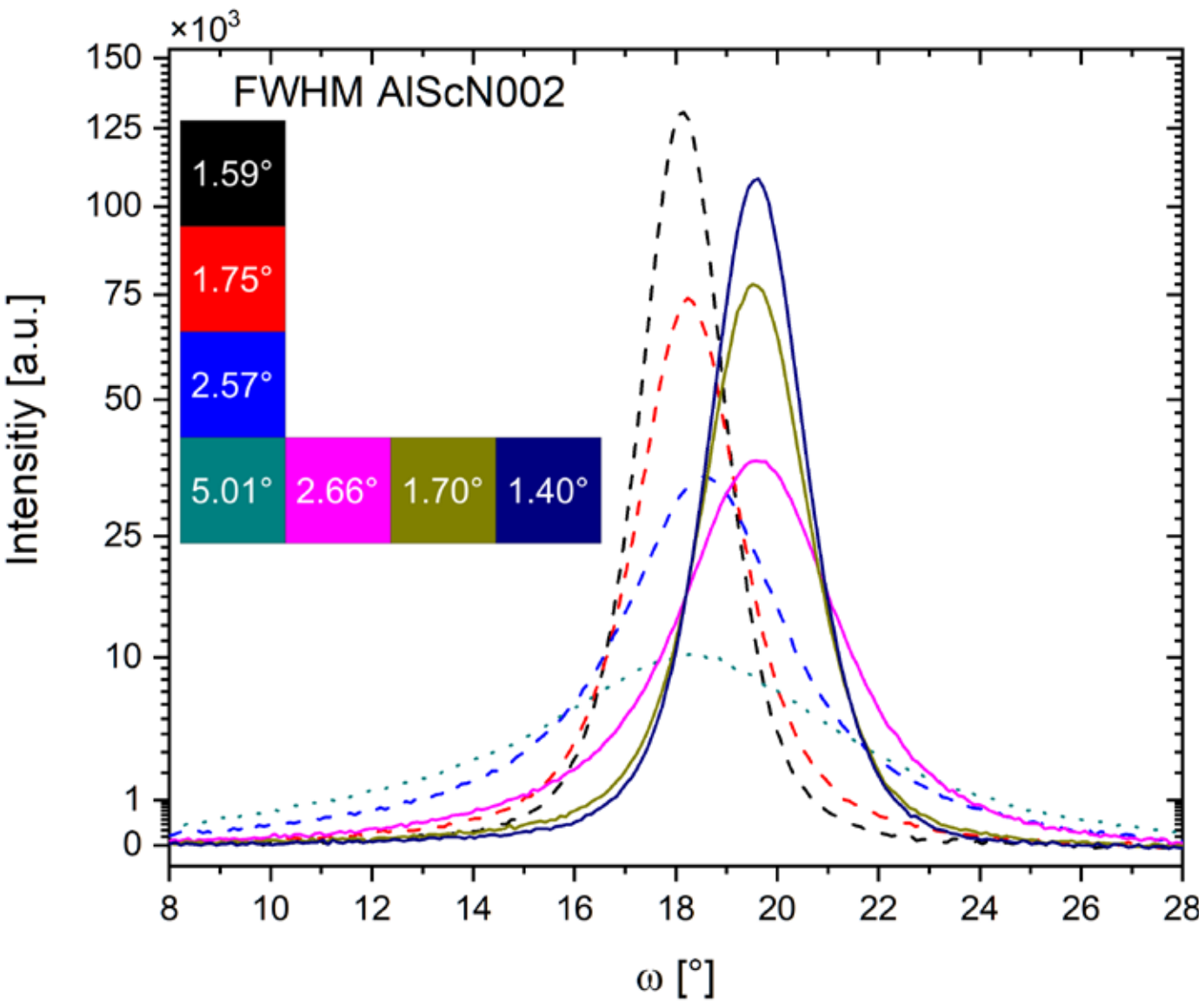

*Figure 6: Rocking curves for the $Al_{0.6}Sc_{0.4}N$ (002) reflex. The extracted FWHM is shown in the inset in the top left. The rocking curves form two groups corresponding to the two measurement directions on the wafer. A clear correlation between FWHM and intensity to the distance from the center can be observed, where the wafer center shows the highest FWHM and the edge the lowest.*

Figure 6 shows that the FWHM and the center angle of the $Al_{1-x}Sc_xN$ (002) reflex are position dependent over the wafer. For the measurement positions one to three only a small shift of the center angle towards lower angles, in comparison to the center position with the index four, is visible while for the positions five to seven a larger shift towards higher angles is notable. The first three positions were measured from the wafer top towards the center, while positions five to seven were measured towards the right side of the wafer. One would expect a rotary symmetry over the wafer, but the difference can be explained by the XRDs beam path geometry. The measurements showing only a small shift are aligned orthogonal to the beam path while the measurements with large shift were parallel. Overall, this allows for determination of the grain tilting. The grains are tilted towards the wafer center and only slightly to the left or right in reference to the wafer surface normal vector.

To fully understand the different center angles for the two measurement directions in the RCs, RSMs around the $Al_{0.6}Sc_{0.4}N$ (002) reflection at the wafer top (Position 1), middle (Position 4) and right side (Position 7) were recorded and are presented in Figure 7. All three maps show the Pt (111), $Al_{0.6}Sc_{0.4}N$ (002) and Mo (110) reflection. The Pt signal at around 40° 2θ is stable for all three positions, as expected from the T2T measurements and RCs. The $Al_{0.6}Sc_{0.4}N$ (002) signal at around 36.5° 2θ behaves differently for the three positions, as already seen in the previous XRD measurements. At position 4 the $Al_{0.6}Sc_{0.4}N$ signal is spread wide in ω and gives overall low intensity, confirming lower texture than on the rest of the wafer.
By comparing the $Al_{0.6}Sc_{0.4}N$ and Mo signal at position 1 and 7 a shift to higher ω angles is visible for position 7, while at position 1 both signals are in line with the Pt signal. Both positions show a compact $Al_{0.6}Sc_{0.4}N$ signal. The RSMs suggest that the $Al_{0.6}Sc_{0.4}N$ growth at the wafer edge occurs slightly tilted towards the wafer center. Together with the geometry of the XRD setup, this implies that the angle between incident beam and the deposited grains varies between positions 1-4 and 4-7. This tilt together with the difference in crystal quality from center to edge can explain the difference in the grain pattern observed in Figure 2.

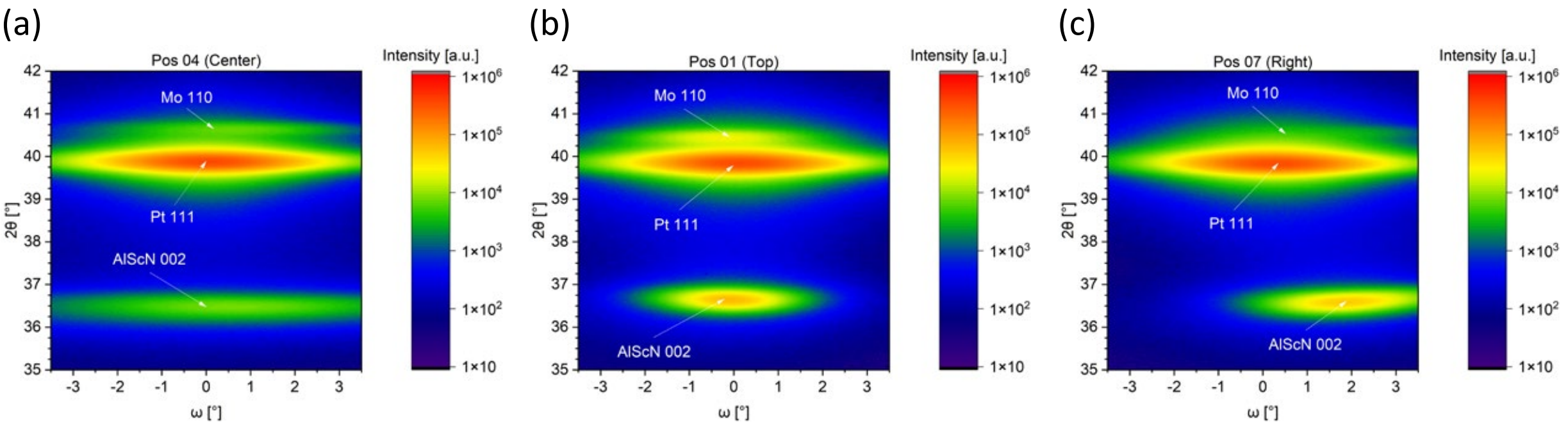

*Figure 7: Reciprocal space mapping of the wafer center (a), wafer top (b) and right side (c). For all three measurement positions, the Pt (111) reflex can be observed at a stable position of 40° 2ϴ, while also a faint reflex from the top Mo electrode at approximately 40.5° 2ϴ is visible. Between a 2ϴ angle of 36° and 37° the $Al_{0.6}Sc_{0.4}N$ (002) is visible for all three measurement positions but with a significant difference between the positions. In the middle a wide reflex is observed, while near the wafer edges narrower responses are observed. Furthermore, a clear shift in ω direction towards larger angles can be observed for the right wafer side in comparison to the wafer top.*

The lattice parameter a was found to be 4.913 Å, while c was determined to be 3.334 Å.

As differences in crystalline quality and (piezo-) electric performance can also depend on film thickness and variations in the chemical composition, we analyzed the distribution of both thickness (Table 1) and composition (Table 2) over the wafer using SEM (EDX).

*Table 1: Film thickness of $Al_{0.6}Sc_{0.4}N$ averaged over three measurements at each indicated position. Thickness uncertainty arises from averaging these three measurements. The measurement was conducted by investigating the freshly cleaved edge of the film via SEM.*

| Position | Thickness [nm] |
|---|---|
| 1 (Top) | 586.5 ± 1.6 % |
| 4 (Center) | 631.2 ± 1.8 % |
| 7 (Right) | 567.2 ± 1.3 % |

The film thickness in Table 1 were recorded on three positions on the wafer: top (Position 1), center (Position 4), and right side (Position 7). The chips were cleaved and evaluated on the newly exposed edge via SEM imaging. Per chip three thickness measurements were performed and averaged together. The thickness uncertainty arises from averaging. A thickness gradient is visible in the recorded values, where in the wafer center the piezoelectric film is thickest and gets thinner towards the wafer edge. The thickness gradient also seems not to be radially symmetrical over the wafer, as change in thickness varies between center to top and center to right side of the wafer.

*Table 2: Elemental composition of the $Al_{1-x}Sc_xN$ film at all seven measurement positions determined via SEM EDX at 8 kV. The oxygen signal in part arises from surface oxidation and the Mo signal from etch residuals of the removed top electrode.*

| Position | Al [at%] | Sc [at%] | N [at%] | O [at%] | Mo [at%] | $Al_xSc_{1-x}N$ composition |
|---|---|---|---|---|---|---|
| 1 (Top) | 29.1 | 20.8 | 47.2 | 3.0 | --- | $Al_{0.582}Sc_{0.418}N$ |
| 2 | 28.9 | 20.7 | 47.5 | 2.8 | --- | $Al_{0.585}Sc_{0.415}N$ |
| 3 | 28.4 | 18.7 | 49.0 | 3.8 | --- | $Al_{0.604}Sc_{0.396}N$ |
| 4 (Center) | 27.7 | 18.1 | 49.3 | 3.5 | 1.4 | $Al_{0.604}Sc_{0.396}N$ |
| 5 | 29.6 | 19.6 | 47.5 | 3.3 | --- | $Al_{0.604}Sc_{0.396}N$ |
| 6 | 29.1 | 19.9 | 47.7 | 3.3 | --- | $Al_{0.595}Sc_{0.404}N$ |
| 7 (Right) | 28.8 | 20.4 | 47.4 | 3.4 | --- | $Al_{0.587}Sc_{0.413}N$ |

The compositional data shown in Table 2 was acquired on the seven positions mentioned in Figure 1. The EDX measurements were performed at an acceleration voltage of 8 kV, on positions where the Mo top electrode was removed during processing. A gradient, similar to the distribution of the film thickness over the wafer is visible in the data. From the wafer center to the wafer edge the Scandium concentration increases, while the Aluminum signal reduces accordingly. This can be explained by the angle and atom-mass dependent sputter yield [27]. The Sc concentration near the wafer edge reaches values higher than the Sc-content of the target material. It has to be noted, that the film composition was determined by EDX while the target composition was evaluated by Inductively Coupled Plasma - Optical Emission Spectrometry (ICP-OES). The Mo signal at position 4 arises from small residues of the top electrodes. High resolution SEM images are available in the supplemental material. The Oxygen signal on all measurement positions can at least in parts be explained by surface oxidation of the exposed film.

The interpolated values for *$d_{33,f}$*, the permittivity $\varepsilon_r$ and the loss tangent *tan(δ)*, in percent, were combined into maps as shown in Figure 8. All three maps show two main regions on the wafer for all parameters. A center region with overall lower values and an edge region with higher values.

The wafer average piezoelectric coefficient is 15.62 pm/V ± 5.3 %. The test structure with the lowest coefficient still showed a *$d_{33,f}$* of 13.66 pm/V while the highest value was recorded at 17.01 pm/V. The average permittivity was found to be 28.32 ± 3.5 %, while the average loss tangent was evaluated to 0.57 % ± 28 %. In absolute values a maximum permittivity of 29.18 was found while the smallest values were evaluated to 25.37. For the loss tangent the extrema are 0.39 % and 1.03 %. All in all, decent homogeneity of the piezoelectric coefficient and relative permittivity was obtained in spite of the increased RC FWHM and lower Sc concentration in the wafer center. Again, further fine tuning of the process will very likely result in even better homogeneity.

a)

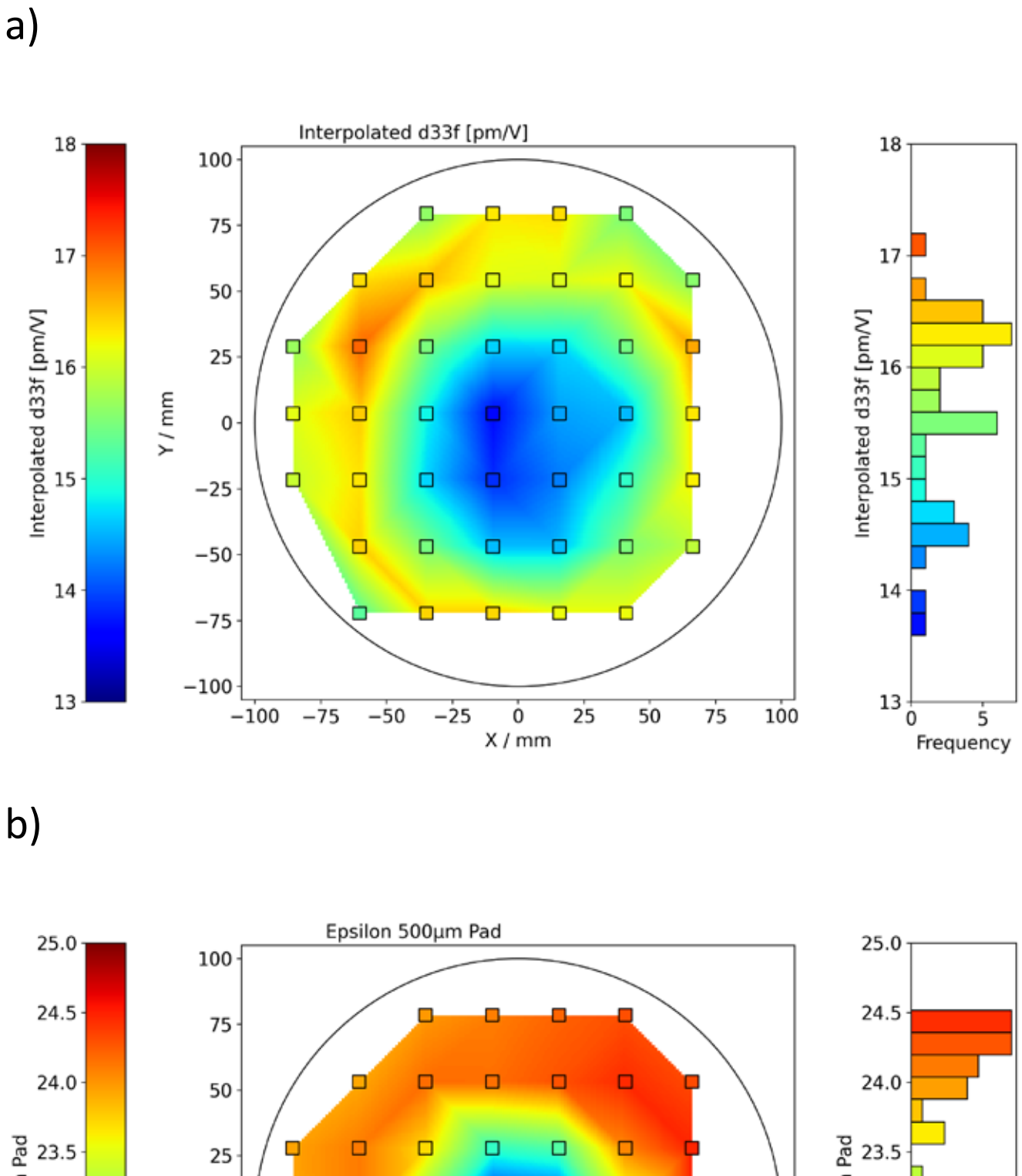

b)

c)

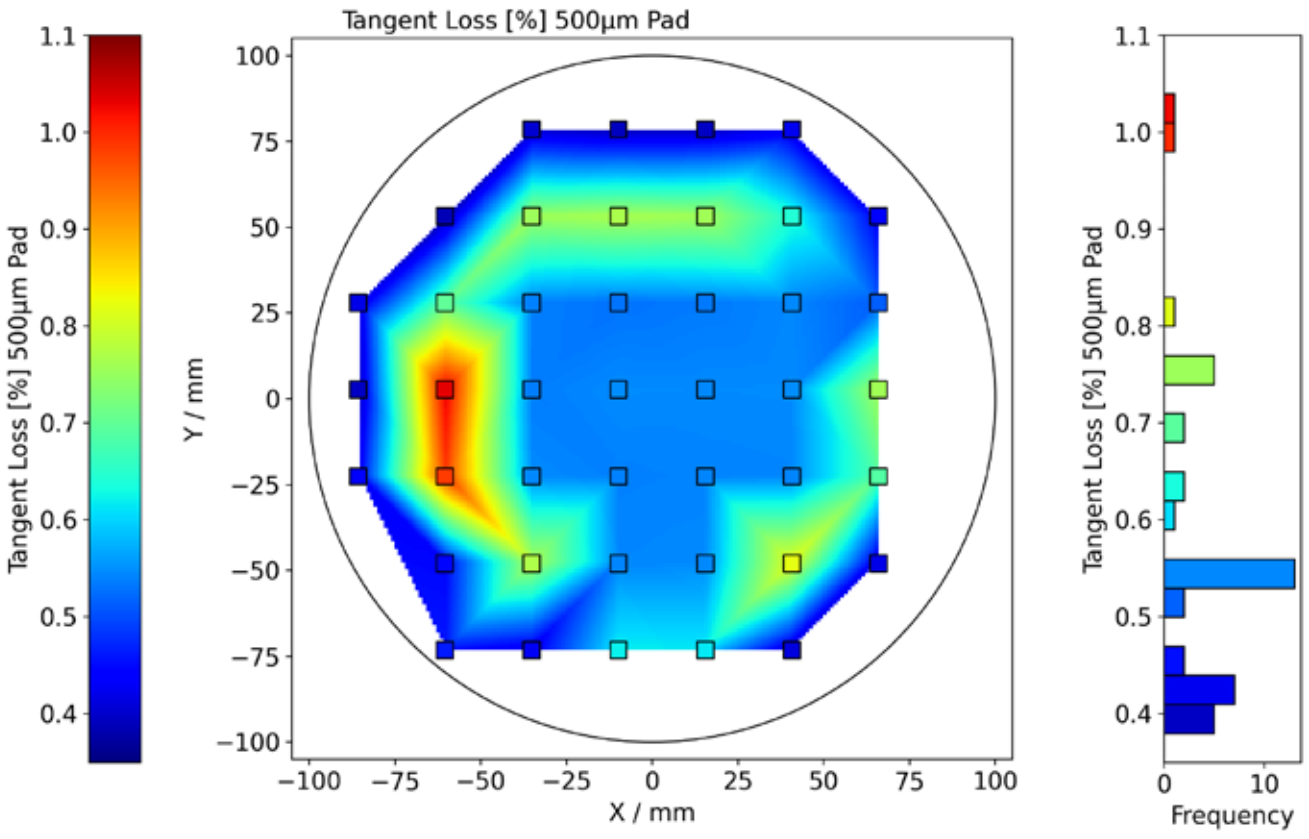

*Figure 8: Wafer maps of (a) interpolated effective out-of-plane piezoelectric coefficient $d_{33,f}$, (b) relative permittivity $\varepsilon_r$ and (c) loss tangent tan(δ). Color in the marked rectangles show measured values while coloring in between arises from interpolation. A clear difference between the wafer center and edge is visible, where albeit a higher loss-factor a better piezoelectric performance is observed. Maps assume a constant film thickness of 595 nm.*

A gradient is observed in the wafer maps for the piezoelectric coefficient and the permittivity. For the piezoelectric coefficient (Figure 8a) the highest values are recorded at the wafer edges with a reduction towards the wafer center. This correlates to the variation in Scandium content over the wafer, as mentioned in Table 2, and the film texture shown via the FWHM of the $Al_{0.6}Sc_{0.4}N$ (002) reflex in Figure 6. This confirms that the piezoelectric coefficient increases with increasing Scandium content of the film and with a higher degree of c-axis oriented grains in the film. For the relative permittivity an even more pronounced radius dependent behavior occurs (Figure 8b). A difference of around 20 % between center and edge is visible. The permittivity is mainly dependent on the film thickness and the Scandium content of the film. As the film thickness only varies by around 5 % (Table 1) we attribute the variation of the permittivity over the wafer predominantly to the change in Sc concentration in the film, as it varies by almost 2 % between center and edge (Table 2). For the loss tangent no gradient was observed, but a ring of high loss factor at a medium radius on the wafer (Figure 8c).

The film was further evaluated for its leakage current under DC conditions, the resulting currents through the film are shown in Figure 9. The leakage follows the same trend on all measured positions. Especially at the highest applied fields a position dependence of the leakage becomes visible, the highest leakage is recorded at the wafer center while near the edge only half the current is recorded. This correlates to the positional trend of the FWHM of the $Al_{0.6}Sc_{0.4}N$ signal and the change in film thickness over the wafer.

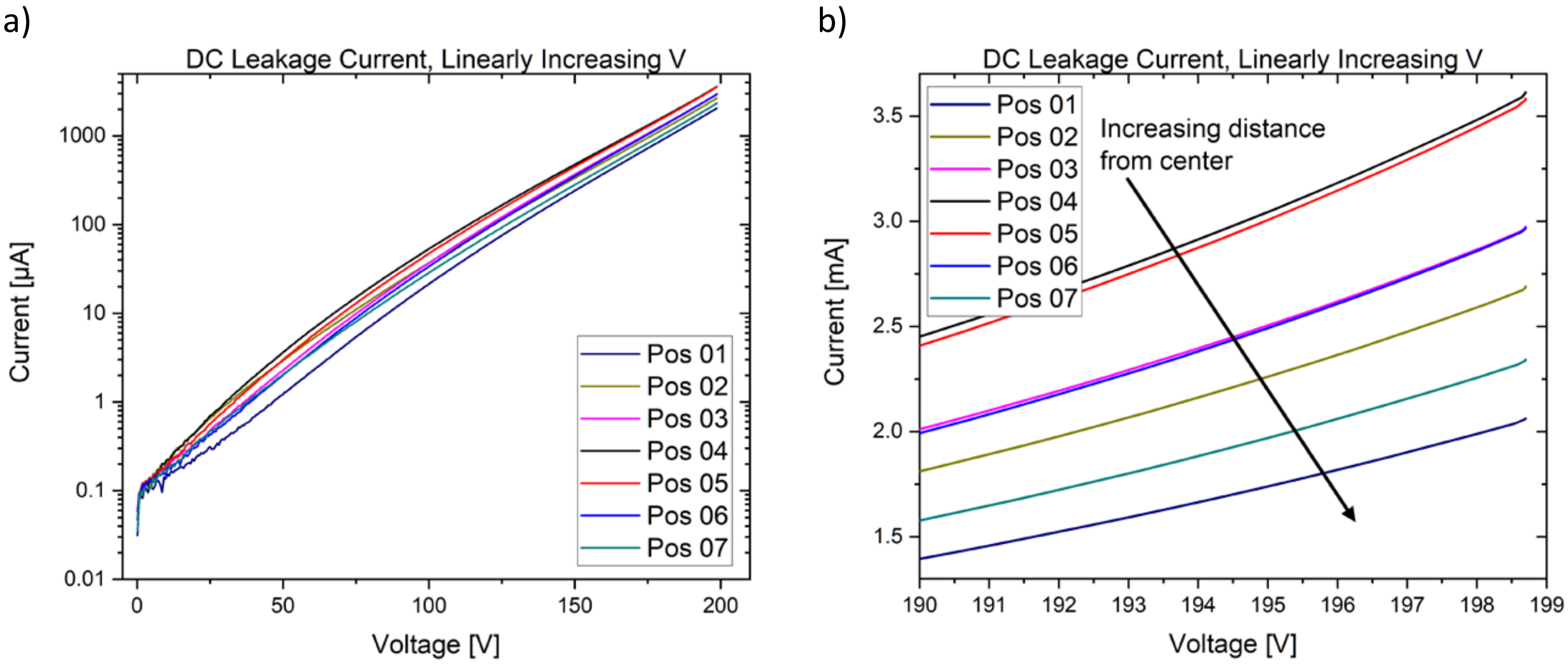

*Figure 9 (a) Logarithmically plotted current trough the AlScN film under linearly increasing DC voltage, (b) image section of (a) at maximum applied voltage with linear scaling.*

BDM measurements revealed an exponential increase of the leakage current with increasing DC voltage. At low voltages (Figure 9a) the seven measured positions show similar behavior to the recorded loss factor map (Figure 8c) whereas the center and edge of the wafer exhibits low leakage while the intermediate radius shows high leakage currents. A direct comparison is not advisable since both measurements were done at different frequency regimes, BDM at DC and CVM measurement at 5 kHz. With increasing voltage this correlation changes as the leakage increases at different rates on the seven positions. At high voltages (Figure 9b) a clear radius dependent leakage behavior is observable. The highest leakage current is recorded at the wafer center and almost halves towards the wafer edge. This correlates to the FWHM of the $Al_{1-x}Sc_xN$ (002) reflex and to some degree to the film thickness. At the same time the lowest DC leakage current is recorded at the positions with the highest Sc content. Thus, we conclude that the DC leakage at high voltages is mainly due to local film texture and grain boundary shape. As a consequence, we expect a decrease in DC leakage currents after further process optimization and thus more homogeneous growth over the wafer as already observed at the wafer edges.

In addition to the piezo- and dielectric properties, the ferroelectric behavior of the film was investigated as well. At all seven measurement positions, clear ferroelectric switching behavior was observed. An exemplary polarization and current loop for position six is shown in Figure 10. The typical square polarization loop of $Al_{0.6}Sc_{0.4}N$ is visible, with coercive fields ($E_c$) at 201 V/µm and -263 V/µm respectively. A remanent polarization of 88.36 µC/cm$^2$ was found. The polarization values result from integration of the measured UI curve.

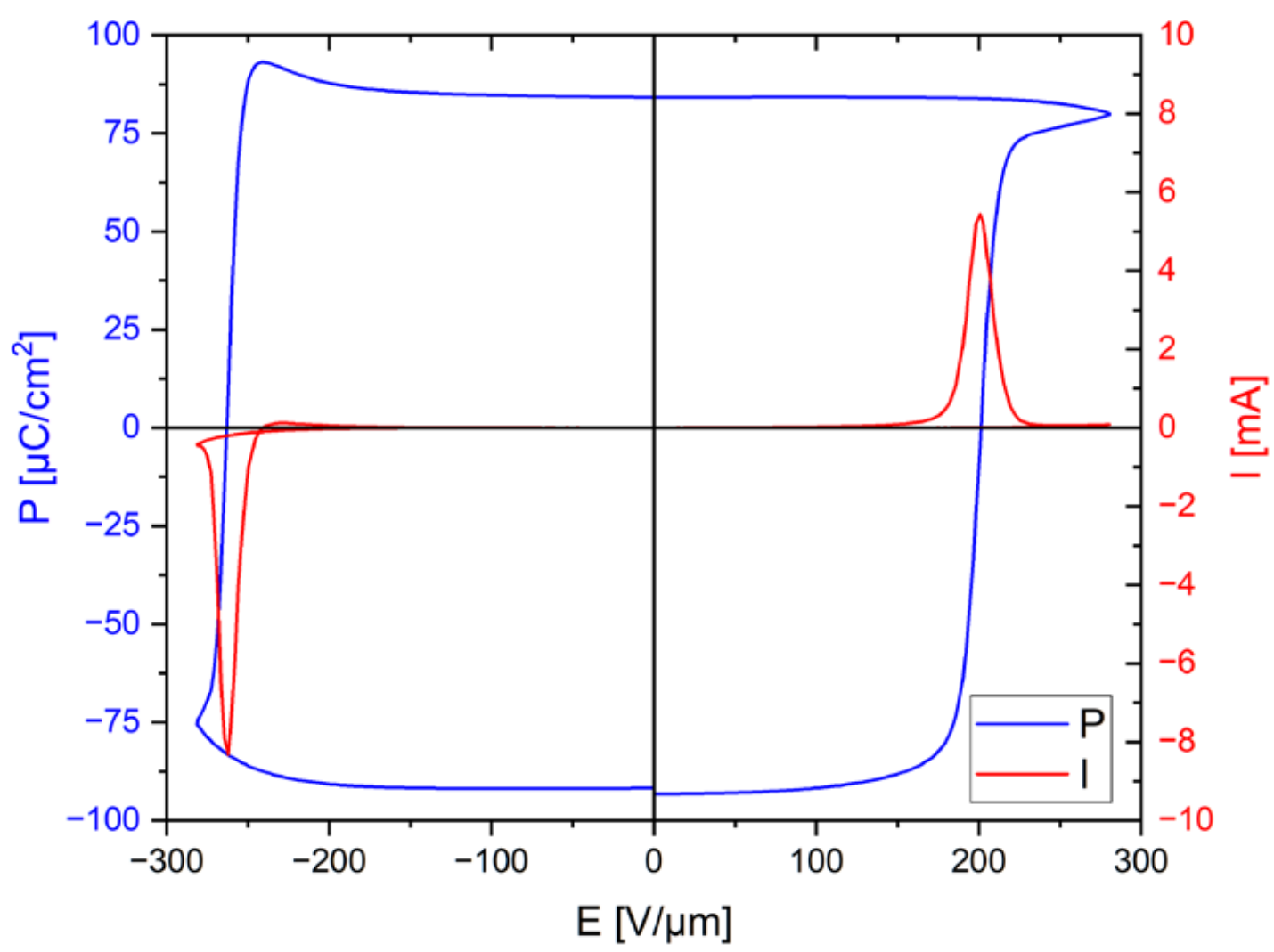

*Figure 10: Polarization (blue) and current (red) loop in dependence of applied electric field for a 500 µm side length square capacitor structure with 595 nm thick $Al_{0.6}Sc_{0.4}N$ and asymmetric electrode materials (Bottom: Pt, Top: Mo). Leakage corrected via PUND scheme.*

Figure 11 shows the IV current loops for all seven position that were characterized. Here also a clear trend in dependence of the wafer radius is visible. The height of the switching spike decreases towards the wafer center while the coercive field increases from average 214 V/µm to average 274 V/µm.

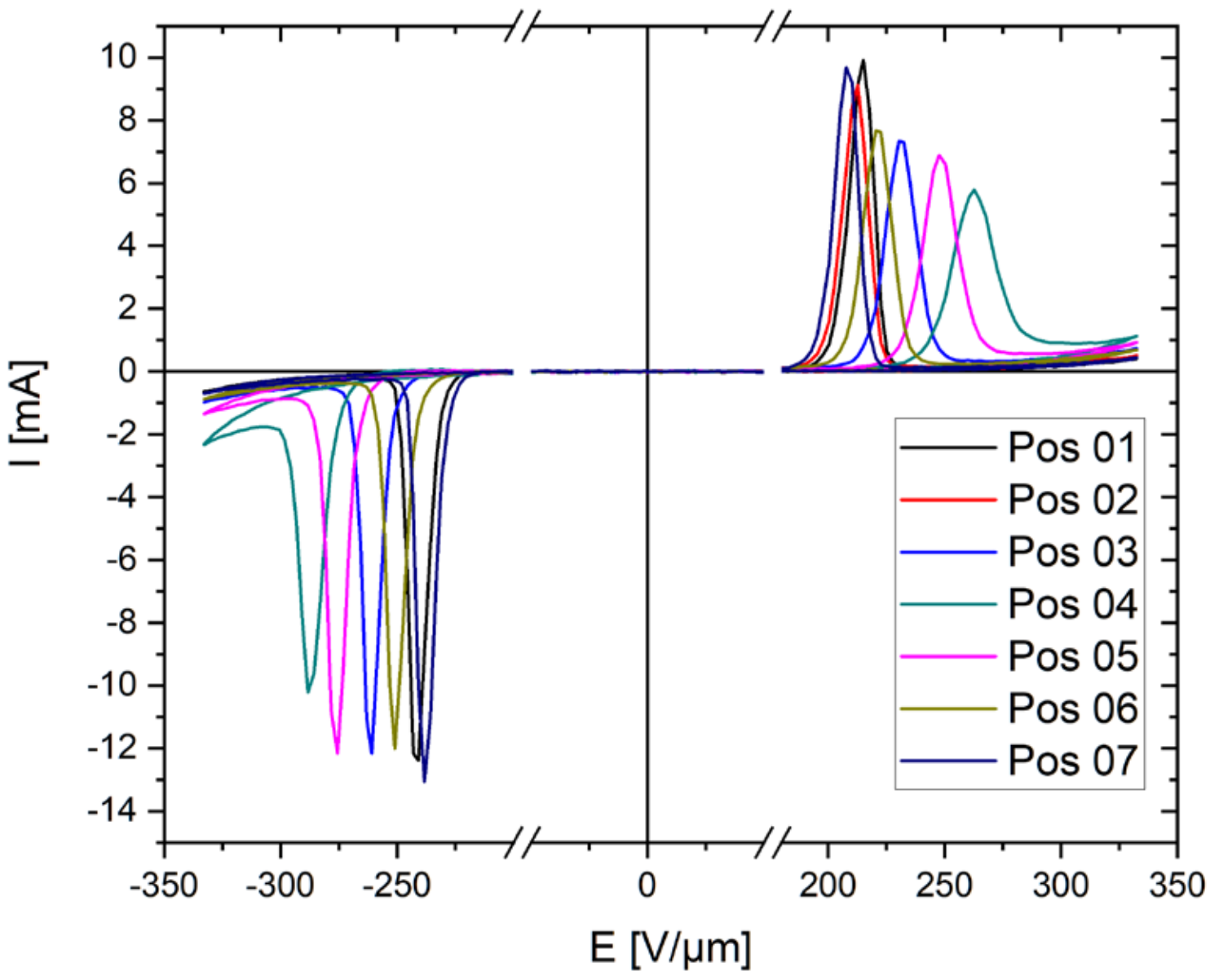

*Figure 11: IV - loops of all seven measurements position acquired via the PUND scheme.*

As shown in Figure 11, a radius dependent ferroelectric switching behavior is observed. For this multiple explanation can be thought of. As seen in Table 2, a gradient in Sc concentration is present in the film. With increasing Sc concentration in the film a decrease in the coercive field is expected, but the variation is only in the range of 2 % and thus cannot completely explain the change [28]. Another explanation is the change in film texture. With worse structure more domain pinning is expected, which would be visible by a decrease in switching current and a widening of the switching peak. Both effects are visible in the measurements near to the less textured wafer center. They have an influence in the measurement but will likely be only one contribution to the change in the coercive field. The other major contribution will come from the variation in film thickness over the range of the wafer. All seven positions assume a fixed thickness of 595 nm for the calculation of the applied electric field, whereas in reality a thickness gradient is present, as shown in Table 1. Due to this variation, the real electric field on each position will differ from what is plotted in Figure 10. Whereas the thickness variation is in the range of 10-15 %, the calculated coercive field varies in the range of 30 %. Since film thickness and coercive field have an approximately linear relation [28, 29], it can explain half of the change. The remaining variation in $E_c$ and the change in peak current and width of the switching peak can likely be attributed to the gradient in film composition and texture.

**Discussion**

We give the first report on $Al_{0.6}Sc_{0.4}N$ films deposited in an industrial tool on full 200 mm wafers, with the help of a 300 mm AlSc target – a setup that can be considered standard for mass production of MEMS devices. In spite of the high Sc content, a low density of AOGs was confirmed over the whole wafer, demonstrating that the c-axis oriented growth that is a necessity for good piezoelectric performance can be sufficiently stabilized under the chosen process conditions. A variation in grain size and shape was none the less apparent and correlates with an improved crystalline texture, with $Al_{0.6}Sc_{0.4}N$ (002) RC FWHMs of around and below 2° achieved over most of the wafer. This confirms that $Al_{0.6}Sc_{0.4}N$ can be grown with very good crystalline properties on large substrates. The remaining suboptimal texture in the wafer center will be treatable through further process optimization. In this deposition campaign, only a small number of wafers was processed in order to get an initial understanding of the general feasibility to deposit films with very high Scandium concentration from a 300 mm AlSc target. Without an extensive parameter variation deposition series, the resulting films of these initial depositions exceeded expectations and were processed for in depth analysis. As reduced crystalline quality does not correlate with an increase in Sc concentration and no signs for the onset of a general phase transition to rock salt was observed in the wafer center, there is no hard limit for the integration of AlScN films with Scandium contents near the phase transition into for example MEMS devices. Through our structural investigations, it can be expected that the improvement of this area will be possible solely through deposition parameter variations. Thus, from a structural point-of-view, the feasibility of fabricating $Al_{0.6}Sc_{0.4}N$ thin films with industry-grade tools and processes was confirmed.

By increasing the Scandium content from 30 % to 40 % in the deposited films a strong increase in piezoelectric performance could be observed. Averaged over the wafer, an effective piezoelectric coefficient $d_{33,f}$ of 15.62 pm/V was found. At the same time the relative permittivity of the films increases moderately to an average of 28.32. Both was expected and matches to the overall trends found when varying the Scandium content in $Al_{1-x}Sc_xN$. While the increase of the piezo coefficient is desired and beneficial for the film performance, the increase in permittivity means larger losses due to reactive power in operation. Even though the increase in permittivity is undesirable, the overall increase in piezo performance overcompensates the disadvantages. In addition, even this enlarged value is still orders of magnitude smaller than the permittivity of other ferroelectric ceramics like PZT which are used in current devices. The strong increase in piezoelectric coefficient is especially interesting for applications with high force demands like MEMS driving of cantilevers and operation out of resonance. Furthermore, the stiffness of the $Al_{1-x}Sc_xN$ film is reduced with increasing Scandium content, improving the driving performance even more. The loss tangent was recorded, averaged over the wafer, to be 0.57 %, increasing in the same manner as the other values. This increase could correlate to the slightly tilted grain orientation over the wafer und thus decreased film texture, therefore further process optimization will most likely reduce it. While a higher loss factor causes more material heating under AC or even resonant driving conditions, this extremely high Scandium content film is, as mentioned before, especially interesting for high force and quasi-static driving scenarios in MEMS there this increase in the loss factor will have no large impact on the device performance.

In Table 3 the wafer average piezo coefficient $d_{33,f}$, relative permittivity $\varepsilon_r$ and loss tangent $tan(\delta)$ of the here presented $Al_{0.6}Sc_{0.4}N$ film are listed in the first row. For comparison also wafer average values found in literature are listed [12–14]. They noted that the resulting film composition contained 3 % more Scandium than the target, this matches with our EDX data, showing higher Scandium concentration in the films than in the metallic targets.

Piezo- and dielectric characterization of the grown films was performed via full wafer mapping on a DBLI for the cited three sources and our films, allowing for rather direct comparison of the values. As not all measurement parameters are given in the literature, differences arising from e.g. different measurement frequencies cannot be accounted for here. In direct comparison to the literature averages the piezoelectric coefficient was increased by 4.4 pm/V or 40 %, the permittivity increased by 10.65 or 60 % while the loss factor rose by 0.24 % in absolute values which is a relative increase of 72 %. It is clear that $Al_{1-x}Sc_xN$ films near the phase transition to the cubic rock salt phase show a large potential for commercial applications. While the undesired increase in the dielectric parameters appear large in relative terms, their absolute values can still be considered small, especially in in comparison to other ferroelectric ceramics in use today.

*Table 3: Wafer average piezo coefficient $d_{33,f}$, relative permittivity $\varepsilon_r$ and loss tangent $tan(\delta)$ for $Al_{0.6}Sc_{0.4}N$ and $Al_{0.7}Sc_{0.3}N$ films measured on a DBLI. Data for 40 % Sc content is this work while data for 30 % is collected from literature.*

| Film Composition | $d_{33,f}$ [pm/V] | $\varepsilon_r$ [1] | $tan(\delta)$ [%] |
|---|---|---|---|
| $Al_{0.6}Sc_{0.4}N$ | 15.62 | 28.32 | 0.57 |
| $Al_{0.7}Sc_{0.3}N$ [12] | 9.5 -11.3 | No Data | 0.2 – 0.6 |
| $Al_{0.67}Sc_{0.33}N$ [13] | 10.9 - 11.8 | 17.5 - 17.8 | 0.35 - 0.37 |
| $Al_{0.67}Sc_{0.33}N$ [14] | 11.8 | 17.7 | 0.21 - 0.25 |

The average AlScN film stress was in the desired slightly tensile regime between 100 MPa and 200 MPa. It could be controlled by same set of parameters as previous depositions runs with a $Al_{0.7}Sc_{0.3}$ alloy target. Full wafer ferroelectric behavior, even in areas with sub optimal growth like the wafer center, shows the films potentials after further process optimization. Potential integration into multi-layer stacks to further multiply the output force, the use in ferroelectric field effect transistors (FeFETs) or ferroelectric random access memory (FeRAM) is interesting due to the higher piezo performance respectively the lower switching fields.

Further work is planned to increase the area of optimal growth to the full wafer by means of process optimizations on Pt electrodes. After progressing on Pt electrodes, other systems will also be considered for process optimization like Mo electrodes and direct deposition on Si or $SiO_2$.

## Conclusion

In conclusion, we developed a reactive sputtering process which is able to growth well c-oriented $Al_{1-x}Sc_xN$ films with highest Scandium content (x = 0.4) from a 300 mm target economically on large areas, namely full 200 mm standard Si wafers in an industrial tool. The whole wafer exhibits very high piezoelectric performance as well as ferroelectric behavior. With the current process parameters and in absolute units the wafer edge showed the highest piezoelectric performance which was correlated with the best growth. This will enable the next generation of piezoMEMS devices with superior displacement, linearity and Back-End-of-Line capability while at the same time keeping power uptake at a minimum. Further plans include an improved deposition process to grow a more homogenous film thickness over the whole wafer and extend the process capabilities to other electrodes like Mo and direct deposition on Si or $SiO_2$. The remaining challenges, especially the wafer in wafer homogeneity, can most likely be overcome with more deposition process optimization as large areas of the wafer show excellent performance.

## Acknowledgment

Parts of this work were founded by the Federal Ministry of Education and Research (BMBF) project "ForMikro – Salsa" (grant number: 16ES1053).

The authors thank all staff members of the ISIT fab contributing to the successful development of the process and the fabrication of the structures. Special thanks are devoted to the process engineers for their skill, advice and patience.

*The authors declare no conflict of interest*.